\PassOptionsToPackage{dvipsnames}{xcolor}
\documentclass[10pt,sigconf,nonacm,screen]{acmart}

\usepackage[all]{nowidow}
\usepackage{xcolor}
\usepackage{subcaption}
\usepackage{hyperref}
\usepackage{acronym}
\usepackage{siunitx}
\usepackage{pdfpages}

\usepackage[capitalise,noabbrev]{cleveref}
\crefformat{section}{\S#2#1#3} 
\crefformat{subsection}{\S#2#1#3}
\crefformat{subsubsection}{\S#2#1#3}

\begin{document}
\includepdf[pages=-, scale=1.0]{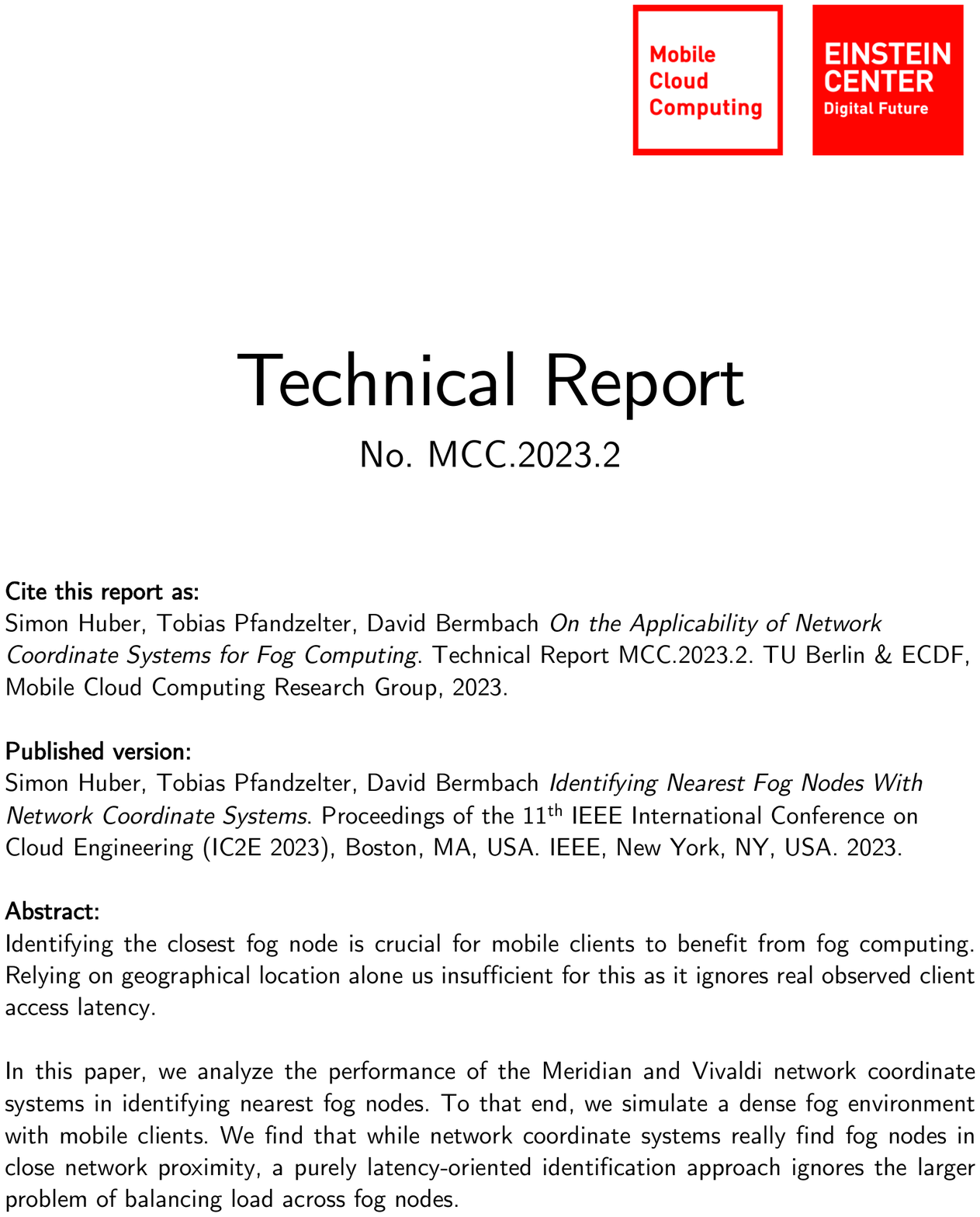}

\author{Simon Huber}
\affiliation{%
    \institution{TU Berlin \& ECDF}
    \city{Berlin}
    \country{Germany}}
\email{shr@mcc.tu-berlin.de}

\author{Tobias Pfandzelter}
\affiliation{%
    \institution{TU Berlin \& ECDF}
    \city{Berlin}
    \country{Germany}}
\email{tp@mcc.tu-berlin.de}

\author{David Bermbach}
\affiliation{%
    \institution{TU Berlin \& ECDF}
    \city{Berlin}
    \country{Germany}}
\email{db@mcc.tu-berlin.de}

\title{On the Applicability of Network Coordinate Systems for Fog Computing}

\begin{abstract}
    Identifying the closest fog node is crucial for mobile clients to benefit from fog computing.
    Relying on geographical location alone us insufficient for this as it ignores real observed client access latency.

    In this paper, we analyze the performance of the Meridian and Vivaldi network coordinate systems in identifying nearest fog nodes.
    To that end, we simulate a dense fog environment with mobile clients.
    We find that while network coordinate systems really find fog nodes in close network proximity, a purely latency-oriented identification approach ignores the larger problem of balancing load across fog nodes.
\end{abstract}

\maketitle

\section{Introduction}
\label{sec:intro}

Fog computing bridges the gap between central cloud data centers and mobile clients with intermediary and edge nodes that provide computing services in close proximity and with low latency~\cite{paper_bermbach2017_fog_vision,bonomi2012fog}.
To benefit from fog computing, clients must be able to identify their nearest node, which is non-trivial in the loosely coupled fog with different access networks~\cite{9946322,pfandzelter2023fred}.
Network-only techniques are insufficient, as the node must also be available and have sufficient capacity.
Similarly, approaches based on geographic location of nodes and clients ignore network characteristics~\cite{Brambilla.2014,paper_hasenburg2020_disgb,Qayyum.2018}.
The discovery and selection of a fog node is a continuous process as fog clients are usually mobile, e.g., connected vehicles or IoT devices, and therefore change their network and physical position~\cite{Garg.2007,paper_bellmann2021_predictive_replication_markov,poster_pfandzelter2021_predictive_replica_placement_poster}.
The naive approach of probing each node in the network will theoretically lead to a perfect result but is not scalable due to the large communication overhead.

There exist some approaches for identifying a nearest node for peer-to-peer (P2P) systems, yet it is unclear if those can directly be applied to fog computing.
In this paper, we aim to close this gap by applying the \emph{Meridian}~\cite{Wong.2005} and \emph{Vivaldi}~\cite{Dabek.2004} \emph{network coordinate systems} to fog computing.
We analyze the performance of both systems in a simulation of mobile clients moving through a distributed fog network.

\section{Network Coordinate Systems}
\label{sec:background:ncs}

Network coordinate systems (NCS) evolved from virtual distance tables with the rapid grow of the Internet~\cite{Guyton.1995,Francis.2001}.
By assigning virtual coordinates to every node in the network, the logical distance could be approximated by triangulation.
Francis et al.~\cite{Francis.2001} proposed \emph{IDMaps} to properly compute these distances by the usage of central landmarks that serve as reference points in the systems.
Two prominent later systems are \emph{Meridian}~\cite{Wong.2005}, which increased accuracy using a multi-resolution ring architecture, and \emph{Vivaldi}~\cite{Dabek.2004}, a decentralized NCS based on a spring-relaxation system rather than central landmarks.

\subsection{Centralized NCS}
\label{sec:ncs:meridian}

In centralized NCS, landmarks serve as reference points for other nodes, promising high stability and accuracy if the landmarks themselves are stable.
A major drawback of this architecture is the higher message overhead and consequently the bottlenecks these landmarks can form.

\begin{figure}
    \centering
    \includegraphics[width=0.5\linewidth]{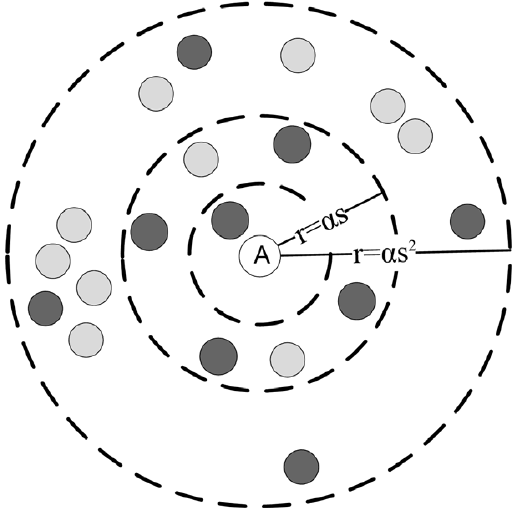}
    \caption{Meridian nodes use non-overlapping rings of exponentially increasing radii to keep track of the distance to other nodes~\cite{Wong.2005}}
    \label{fig:meridianrings}
\end{figure}

Meridian is a centralized NCS that proposes a direct approach for nearest node discovery.
Meridian nodes structure the surrounding nodes of the network into rings depending on their latency, as shown in \cref{fig:meridianrings}.
These rings are concentric, non-overlapping, and have increasing radii.
The $i$-th ring has an inner radius of $r_i =\alpha s^{i-1}$ and an outer radius of $R_i =\alpha s^{i}$ for $i>0$, where the ring base $\alpha$ is a constant as well as the multiplicative increase factor $s$.
The innermost ring has an inner radius of $r_0 =0$ and an outer radius of $R_0 = \alpha$.
Nodes are assigned to their ring by measuring the latency/distance $d_j$ and placing them into the $i$-th ring with $r_i < d_j \le R_i$.

Each meridian node has two ring sets identical in size and radii, a primary and a secondary ring set.
The primary ring set keeps track of $k$ nodes per ring, whereas the secondary ring set keeps track of $l$ nodes per ring.
Meridian node performs its queries on the primary ring set, therefore it is critical to keep a sufficient number of nodes within these rings that are optimally geographically distributed in close and remote regions of the network.
This provides a higher utility for the recursive query to find the nearest node to the target.
The secondary ring is filled with substitutes for the primary ring.

Periodically the ring membership for every $k+l$ nodes is reassessed to ensure the highest diversity within the primary rings.
The assessment is done by comparing the usage of the virtual coordinate of each node and keeping the best $k$ members in the primary ring and the others in the secondary ring.
The virtual coordinate tuple of the node $n$ consists of the distances to the other nodes in the same ring $\langle d^{n}_{1},d^{n}_{2},...,d^{n}_{k+l} \rangle$.
The node collects all tuples of its ring members into a latency matrix and iteratively determines which member is the least valuable.
This is done by iterating over all the members and calculating the hypervolume of the $k+l$-dimensional polytope of the matrix without the member.
If the hypervolume changes least without a member, this member is the least valuable, removed from the matrix, and pushed into the secondary ring.
This process is performed until there are only $k$ members left in the latency matrix, which are the new primary ring members.
The virtual coordinates are shared within the network via a gossip protocol, which allows handles altering latencies as well as dropped out nodes in the network.

\begin{figure}
    \centering
    \includegraphics[width=\linewidth]{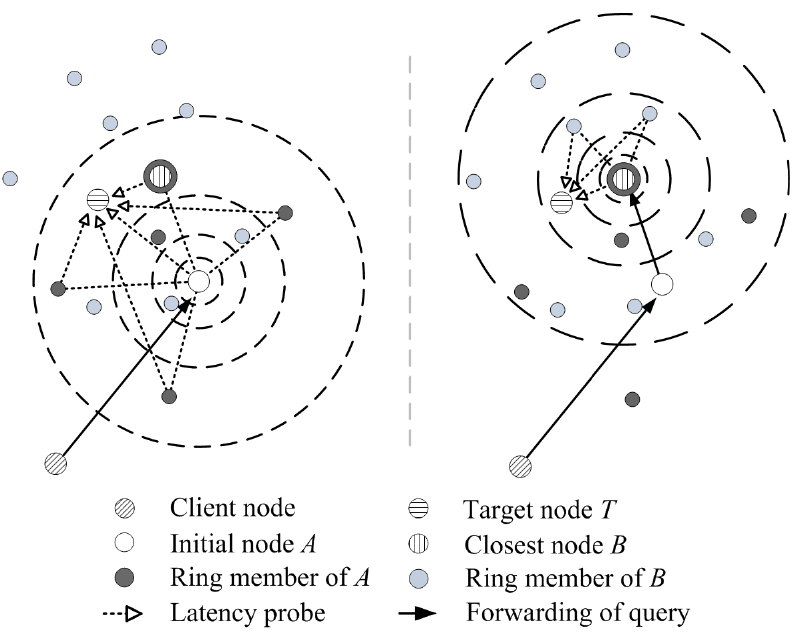}
    \caption{Clients can request closest nodes from any Meridian node \emph{A}, prompting the iterative search~\cite{Wong.2005}}
    \label{fig:meridiansearch}
\end{figure}

We show the nearest node discovery of the meridian framework in \cref{fig:meridiansearch}.
When a client in the systems requests a Meridian node for the nearest node, the meridian node issues every node within a distance of $(1-\beta) \times d$ and $(1+\beta) \times d$ to measure its network distance to the client.
Here, $d$ is the distance between the meridian node and the client and $\beta$ is the acceptance threshold $0 \le \beta < 1$.
Nodes that take more than $(2\beta + 1) \times d$ to answer cannot be closer to the target and are ignored.
The client request is then forwarded to the nearest node from this query, which repeats this process.
If no node answers within threshold of $(2\beta + 1) \times d$, the selected Meridian node is the nearest to the target.
This approach exponentially reduces the distance with every recursion between the node issuing the query and the target until no nodes are closer than the issuing node.

\subsection{Decentralized NCS}
\label{sec:ncs:vivaldi}

Decentralized NCS do not require landmark nodes, decreasing the message overhead and bottlenecks in the system.
But as there are no fixed reference points, the coordinates can oscillate and a higher fault tolerance has to be established in the NCS, which can both negatively influence precision.

Vivaldi is a decentralized NCS that is widely deployed in Internet systems~\cite{YangChen.2011}.
The underlying concept of Vivaldi is a network of nodes connected as in a mass-spring system, where each edge is a spring with a force.
Minimizing the energy of all the springs sends the system into an equilibrium with the length of the springs depicting the virtual distances between the nodes.
This minimization process is also called spring relaxation problem.
It has the following error term to be minimized with $\|x_i-x_j\|$ as distance between the coordinates of node $i$ and node $j$ and the $L_{ij}$ the actual round-trip time (RTT) between the two nodes:

\begin{equation}
    E=\sum\limits_{i} \sum\limits_{j} (L_{ij}-\|x_i-x_j\|)^2
\end{equation}

This squared error term represents the displacement of the spring system.
The rest length of the spring between the nodes $i$ and $j$ is the RTT $L_{ij}$ between them.
The energy of the spring is proportional to the square of the displacement from its rest length.
The force of the spring between the nodes $i$ and $j$ is defined by the energy of the spring times the direction of the spring:

\begin{equation}
    \label{eq:vivaldi_force}
    F_{ij} = L_{ij}- \|x_i-x_j\| \times u(x_i-x_j)
\end{equation}

The net force of the node $i$ is therefore defined as the sum of all forces between the node $i$ and all other nodes:

\begin{equation}
    F_i = \sum\limits_{j\neq i} F_{ij}
\end{equation}

In addition to the length of the spring, the Vivaldi framework implements a positive height vector $h$, which models further packet transmission delays, e.g., queuing delay.
A packet from node $i$ to node $j$ must first travel the height of node $i$, then the length of the spring between $i$ and $j$, and lastly the height of the node $j$.
Therefore, the error induced distance between the two nodes $i$ and $j$ is:

\begin{equation}
    \label{eq:vivaldi_distance}
    d_{ij} = x_i - x_j + h_i + h_j
\end{equation}

Each node in the system has its own virtual coordinate and updated periodically by measuring the RTT to other nodes and recomputing the forces.
It this moves its coordinate $x_i$ in the direction of the forces and eventually converges to a stable coordinate.
As this happens iteratively with every connection, the framework provides an update rule for the coordinate $x_i$ with an adaptive timestep $\delta$ that defines how much the current coordinate of the node is altered by the update:

\begin{equation}
    x_i = x_i + \delta \times L_{ij}- \|x_i-x_j\| \times u(x_i-x_j)
\end{equation}

The adaptive timestep $\delta$ is calculated by the constant fraction $c_c < 1$ times the relative local error given by the local error $e_l$ and the remote error $e_r$:

\begin{equation}
    \delta = c_c \times \dfrac{e_l}{e_l + e_r}
\end{equation}

This timestep makes the coordinate update resilient against high error nodes while providing quick convergence and low oscillation of the network.

When a client requests its nearest node from a Vivaldi node, the Vivaldi node estimates the RTT from the client to all other nodes by calculating the virtual distance.
This can be done without any further message overhead and is a computation with a linear complexity only dependent on the amount of nodes in the network.
For this to work, however, the clients also require a virtual coordinate that they update with the same update rule as the Vivaldi nodes.

\section{Evaluation}
\label{sec:evaluation}

We evaluate the feasibility of NCS-based nearest node discovery in fog platforms with a simulation.

\subsection{Simulation}
\label{sec:evaluation:simulation}

Our simulation scenario is a fog network that mobile clients in the city of Berlin use.
Our simulation uses the \emph{SimPy}~\cite{SimPy.2020} discrete-event simulator.
Ongoing actions such as movements and communication of each entity are designed as processes within the simulation framework.

\subsubsection*{Tested Algorithms}
\label{sec:evaluation:algorithms}

We compare the Meridian and Vivaldi NCS with a baseline and a random approach.
The baseline is the theoretical optimal node selection based on the omniscient simulation environment.
While not attainable in practice, this allows us to compare the overhead and imprecision of the NCS.
For comparison, in the random approach the clients randomly selects a node in the network.

\subsubsection*{Scenario}
\label{sec:evaluation:scenario}

We simulate a \SI{2,25}{\kilo\metre\squared} area in Berlin for 10 minutes simulation time.
Clients in our simulation follow predefined movement patterns obtained from the \emph{Open Berlin Scenario}~\cite{ZIEMKE2019870}.
Clients periodically send tasks to the fog platform, measuring response latency.
Further, clients also respond to latency measurements by nodes to comply with requirements of the NCS protocols.

\begin{figure}
    \centering
    \includegraphics[width=\linewidth]{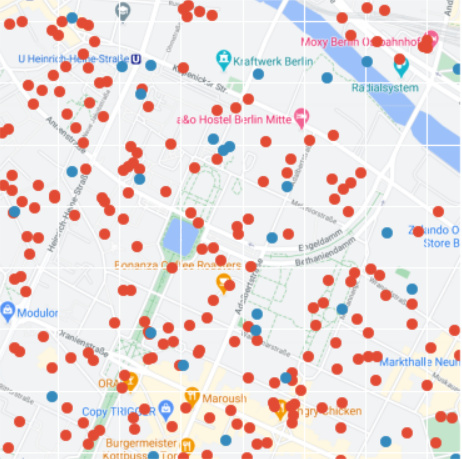}
    \caption{Simulation are in Berlin with 29 fog nodes (blue) and 200 clients (red)}
    \label{fig:simulationarea}
\end{figure}

Each fog node in the network has a dedicated, static geographical location.
Node locations are based on cell tower locations in Germany~\cite{Bundesnetzagentur.2020}.
To abstract from heterogeneous hardware resources of the fog, each node has a predefined number of ``slots'', i.e., concurrent client tasks it can fulfill.
Nodes implement the NCS under test, e.g., serving as Meridian or Vivaldi nodes.
Based on cell tower location data, there are 29 fog nodes in this area.
The number of client slots for a node are assigned randomly, with a total of 318 available slots.
We show an overview of the simulated fog network with fog nodes and clients in \cref{fig:simulationarea}.

\subsubsection*{Latency Approximation}
\label{sec:evaluation:approximation}

Our simulation includes an approximated communication latency between parties.
The latency between participants $i$ and $j$ in a network is made of the sum of the transmission delay, propagation delay, processing delay, and the queuing delay~\cite{Burk.2018}:

\begin{equation}
    L_{ij} = D_{\text{Trans}} + D_{\text{Prop}} + D_{\text{Proc}} +D_{\text{Queue}}
\end{equation}

The transmission delay $D_{\text{Trans}}$ is approximated by a linear regression from the values given by Burk and Lemberg~\cite{Burk.2018} with the bandwidths $B_{i}$ and $B_{j}$ of the nodes \emph{i} and \emph{j}:

\begin{equation}
    D_{\text{Trans}} = -0.008 * min(B_{i}, B_{j}) + 0.088
\end{equation}

The bandwidth $B_i$ of the node $i$ is approximated in this simulation by linearly reducing the available bandwidth of the node with every incoming client to be served, assuming a certain minimum reserved bandwidth for each client.
We assume further that the maximum bandwidth a node can establish to a client is 1Gbps~\cite{Burk.2018}.
Therefore, the bandwidth is calculated by the complement of the ratio of clients the node currently serves $c_i$ and the total amount of slots of the node $s_i$ multiplied by the complement of the service level agreement $\text{sla}$.
This formula is then bounded between $[\text{sla}, 1]$ (in Gbps):

\begin{equation}
    B_{i} = min(1, max(sla, 1 - (1-sla)* \frac{c_i}{s_i}))
\end{equation}

The propagation delay $D_{\text{Prop}}$ is approximated in milliseconds by the sum of the distance $d_{i}$ between the participant $i$ and the nearest cell tower in meters and the distance $d_{j}$ between the participant $j$ and the nearest cell tower in meters multiplied by the propagation speed of the medium in milliseconds per kilometer $v$~\cite{paper_hasenburg2020_disgb}:

\begin{equation}
    D_{\text{Prop}} = (d_{i} + d_{j}) * v
\end{equation}

The processing delay $D_{\text{Proc}}$ is approximated in milliseconds by the hardware factor of the fog node $h$ times the delay factor $c$ in ms plus a static network error $E_{\text{Network}}$ in ms

\begin{equation}
    D_{\text{Proc}} = h * c + E_{\text{Network}}
\end{equation}

The queuing delay $D_{\text{Queue}}$ is approximated by the minimum of
the maximum quality of service error $E_{\text{QoS}}$ in milliseconds and the approximated queue delay in milliseconds taking the bandwidths $B_{i}$ and $B_{j}$ of the participants \emph{i} and \emph{j} into account:

\begin{equation}
    D_{\text{Queue}} = \min(E_{\text{QoS}}, (2 \min(B_{i}, B_{j}))^{-1})
\end{equation}

\subsubsection*{Metrics}
\label{sec:evaluation:metrics}

We collect the mean messages, the mean lost messages, and the mean reconnection requests per client.
For fog nodes we collect the mean incoming messages, mean outgoing messages, and mean total messages.

Further, we collect the connection error $\bar{E}_C$ as average error between the distance between client and selected fog node and distance between client and its optimal fog node.
This metric is collected as the mean root-mean-square-error (RMSE) for every client $i$ and every message $j$ of the client with the total amount of clients $n$, the total amount of messages for the $i$-th client $m_i$, the actual RTT for the $j$-th message of the $i$-th client $d_{ij}$, and the optimal RTT of the same message $f_{ij}$:

\begin{equation}
    \label{eq:Connection Error RSME}
    \bar{E}_C = \frac{1}{n}\sum_{i=1}^{n}{\sqrt{\frac{1}{m_i}\sum_{j=1}^{m_i}{(d_{ij} - f_{ij})^2}}}
\end{equation}

We further measure the error of node service selection $E_S$.
We calculate the mean RMSE of latency between requesting client and selected fog node and client and optimal fog node.

Finally, we measure the mean number of unique selections per discovery timestep.
This illustrates the total number of fog nodes suggested to clients within a timeframe.
A discovery timestep is a timeframe of one second of simulation time within which node discovery is performed by the fog network.

\subsubsection*{Parameters}
\label{sec:evaluation:parameters}

A service level agreement of \SI{0.05}{Gbps} is used to ensure a sufficient bandwidth for each client.
The propagation speed, delay factor, and network error are based on those reported by Burk and Lemberg~\cite{Burk.2018}.
We set the maximum latency acceptable in the fog network at \SI{50}{\ms}.
To observe the impact of client load on NCS behavior, we vary the ratio between total available slots and clients between $0.1$ and $1.0$.
Clients have an initial random start-up delay between 3 and 10 seconds to introduce the clients gradually into the system.
Each client connects to the fog platform periodically within a timeframe of 0.5 and 1 seconds.
The latency threshold is \SI{5}{\ms}, the round trip threshold is \SI{10}{\ms}, and the timeout threshold is \SI{100}{\ms}.

We use suggested parameters for the Meridian NCS~\cite{Wong.2005}.
There are eight rings per ring set, with $k = \lfloor log_{1.6} N \rfloor$ primary members per ring $k$, where $N$ is the number of fog nodes.
The amount of secondary members $l$ is set to $l = N - k$.
The radius of the rings is adjusted to fit the latencies prevalent in the fog platform and has an inner ring radius of $r_i = \alpha s^{i-1}$ and an outer radius of $R_i =\alpha s^{i}$ with the ring base $\alpha = 1$ and the ring factor $s = 1.5$.
The acceptance threshold is set to $\beta = 0.5$.
The ring membership management is performed by each node every \SI{30}{\s}.

The Vivaldi coordinate for each node starts with an initial position in the origin and a local error $e_l$ = 10.
The constant fraction $c_c$ to calculate the adaptive timestep $\delta$ is set to $0.25$.

\subsection{Results}
\label{sec:analysis}

\subsubsection*{Message Delivery}
\label{sec:analysis:messagedelivery}

\begin{figure*}
    \begin{subfigure}{0.32\linewidth}
        \centering
        \includegraphics[width=\linewidth]{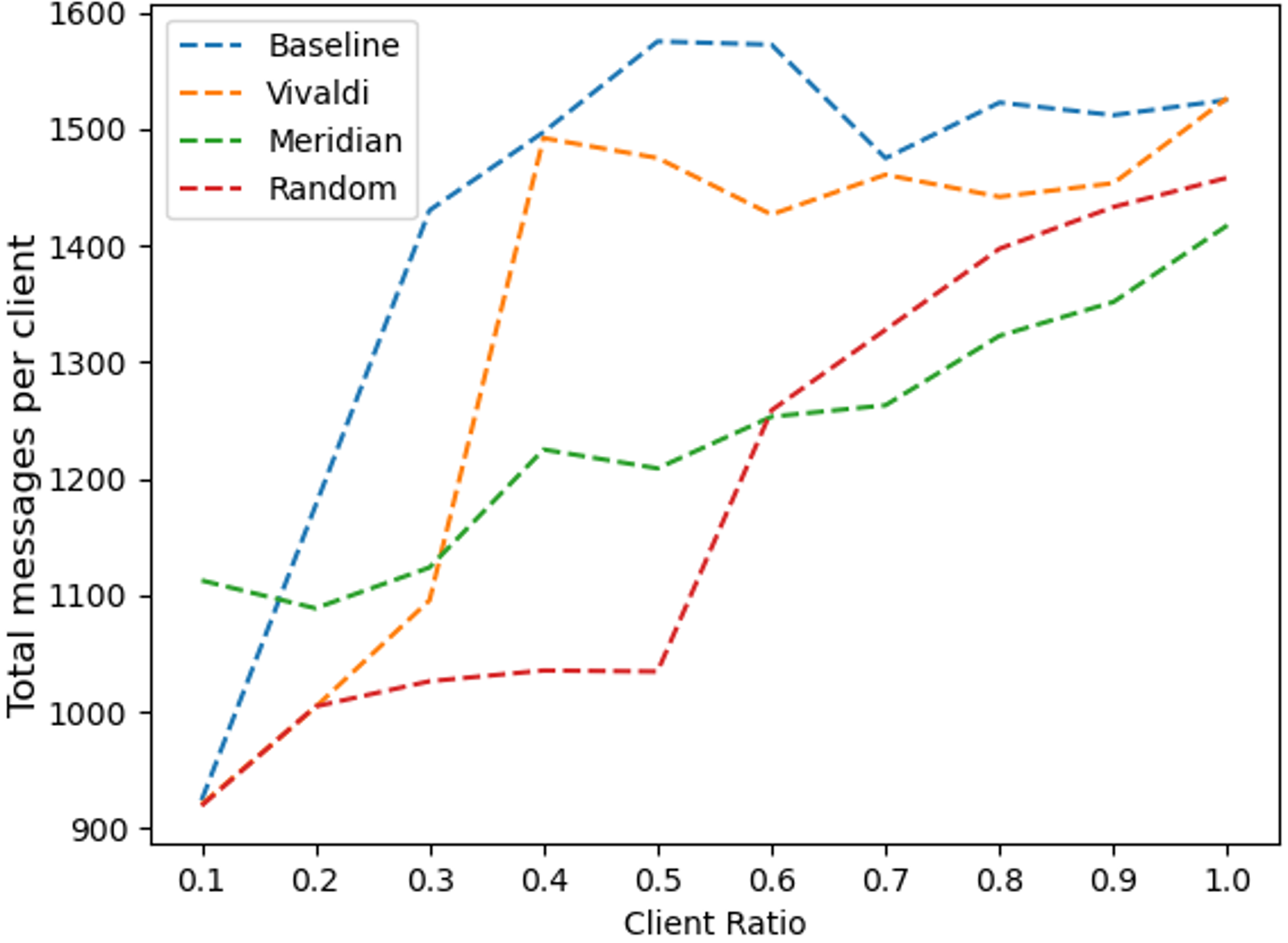}
        \subcaption{Total Messages}
        \label{fig:clientmessages:total}
    \end{subfigure}
    \hfill
    \begin{subfigure}{0.32\linewidth}
        \centering
        \includegraphics[width=\linewidth]{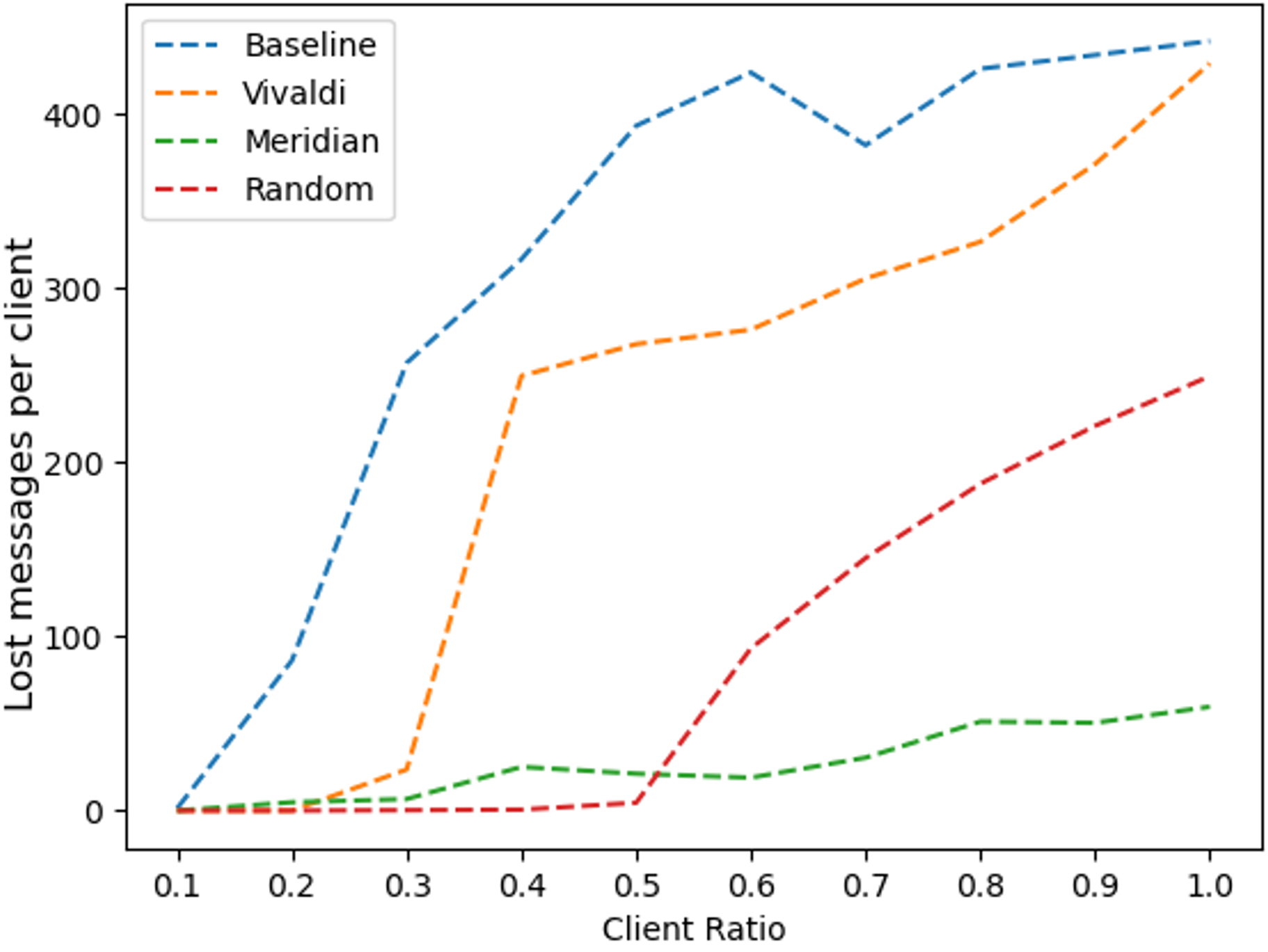}
        \subcaption{Lost Messages}
        \label{fig:clientmessages:lost}
    \end{subfigure}
    \hfill
    \begin{subfigure}{0.32\linewidth}
        \centering
        \includegraphics[width=\linewidth]{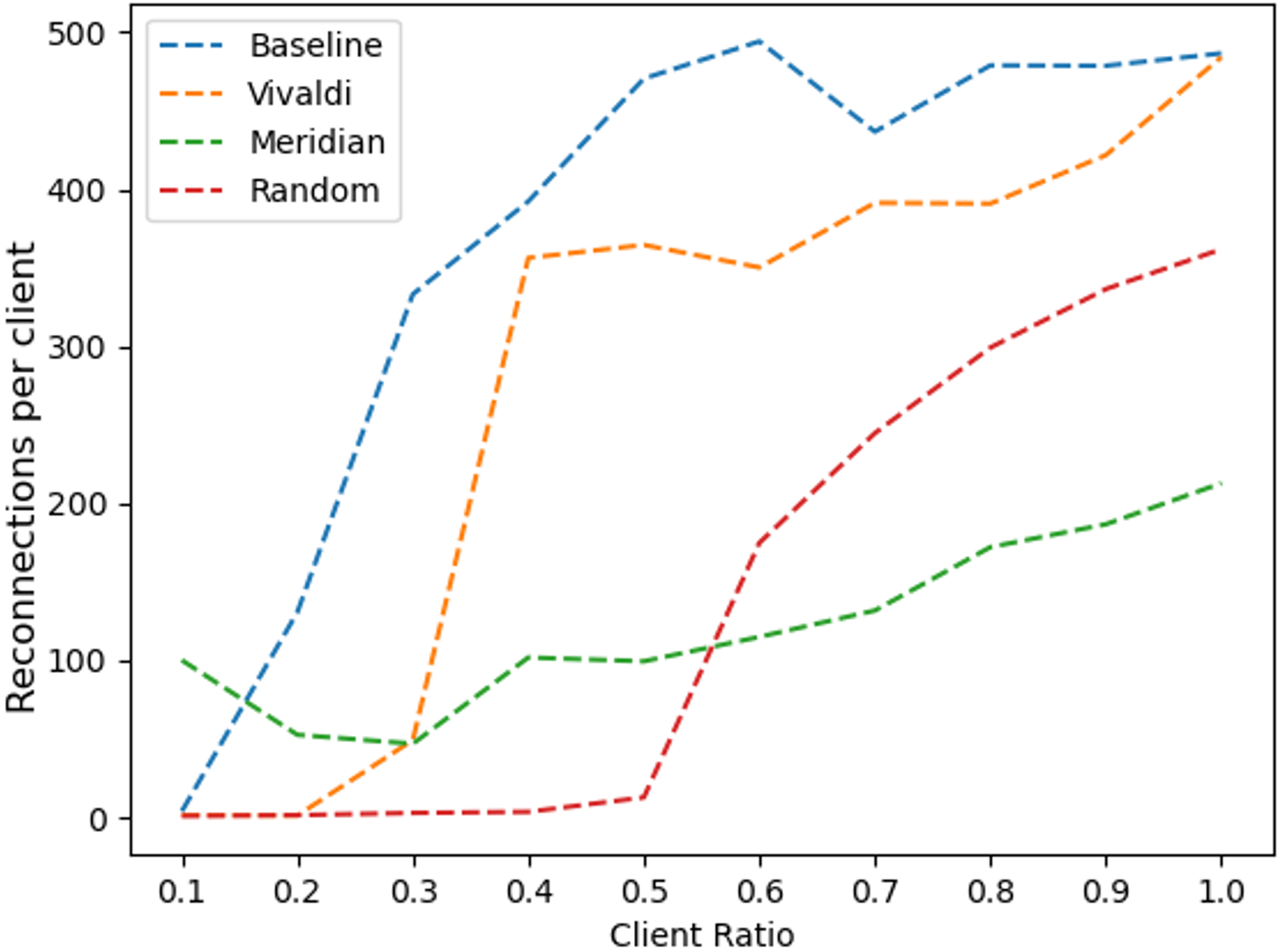}
        \subcaption{Reconnection Requests}
        \label{fig:clientmessages:reconnectionrequests}
    \end{subfigure}
    \caption{Number of messages handled per client for each of the tested algorithms. The total number of messages (\cref{fig:clientmessages:total}) grows with higher client ratios, with baseline requiring the most messages. With increasing client ratios, the number of \emph{lost} messages per client (\cref{fig:clientmessages:lost}) and number of reconnections per client (\cref{fig:clientmessages:reconnectionrequests}) also grows for all tested algorithms, with the Meridian approach requiring the least messages at higher ratios.}
    \label{fig:clientmessages}
\end{figure*}

We show the total number of messages sent per client in \cref{fig:clientmessages:total}.
We observe an increasing amount of messages with increasing client ratio, likely caused by nodes not being able to accept client requests.
We also see that Meridian has a higher number of messages with low client ratios but is more scalable than Vivaldi and our baseline at higher ratios.

For Vivaldi and the baseline we also see a higher number of lost client messages in \cref{fig:clientmessages:lost}.
Messages are lost from the perspective of a client when the fog node is overloaded and cannot process more tasks.
As a result of unprocessed tasks, we also observe higher reconnections for clients, as shown in \cref{fig:clientmessages:reconnectionrequests}.

\begin{figure}
    \centering
    \includegraphics[width=\linewidth]{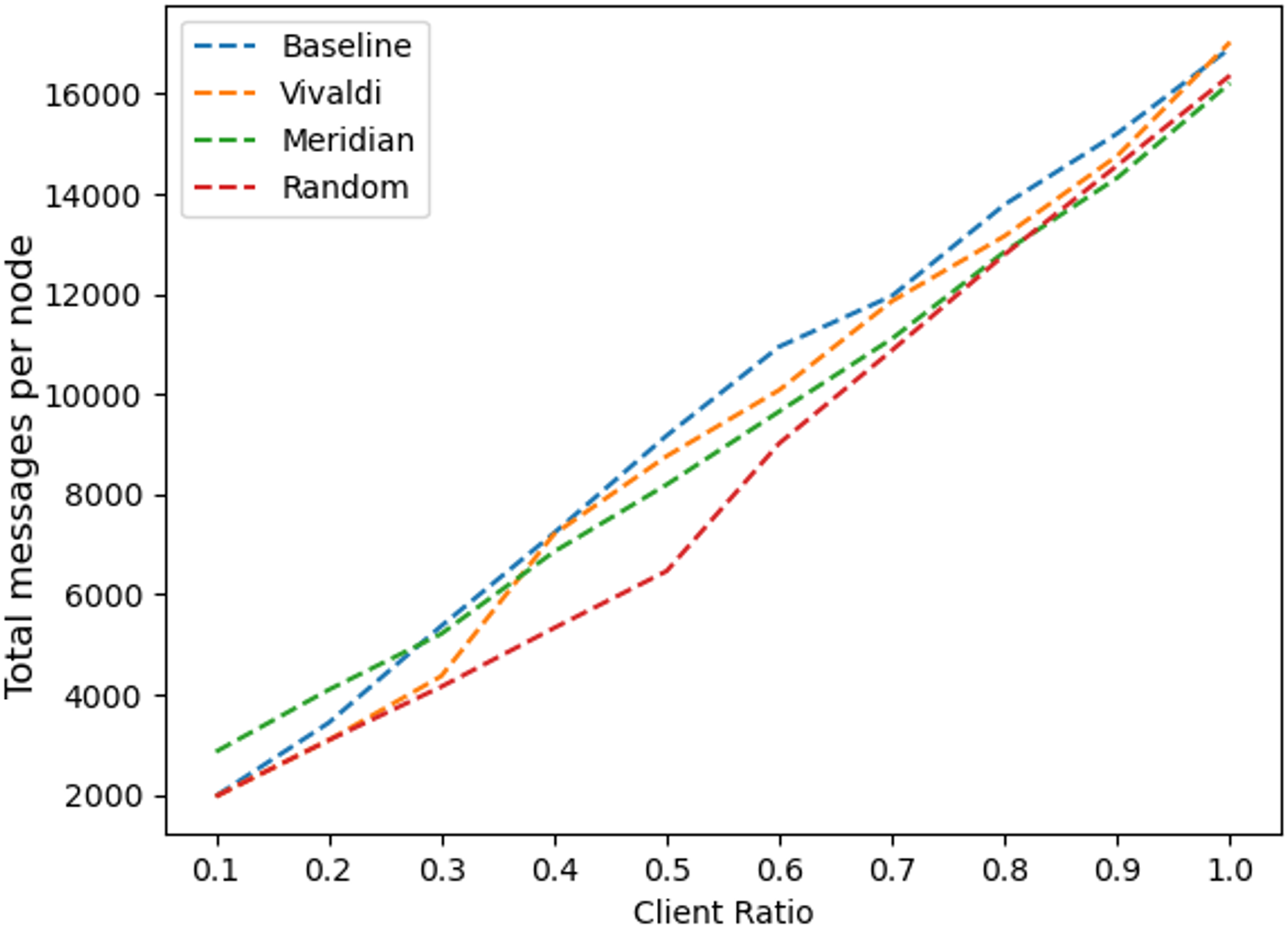}
    \caption{The total number of messages handled per fog node grows linearly with the client ratio for all tested methods.}
    \label{fig:nodemessages:total}
\end{figure}

We show the message load from the perspective of fog nodes in \cref{fig:nodemessages:total}.
We observe a linear growth of the total number of messages with increasing client ratio.
Nevertheless, this effect is seen across all selections strategies.

\subsubsection*{Service Selection}
\label{sec:analysis:serviceselection}

\begin{figure*}
    \begin{subfigure}{0.32\linewidth}
        \centering
        \includegraphics[width=\linewidth]{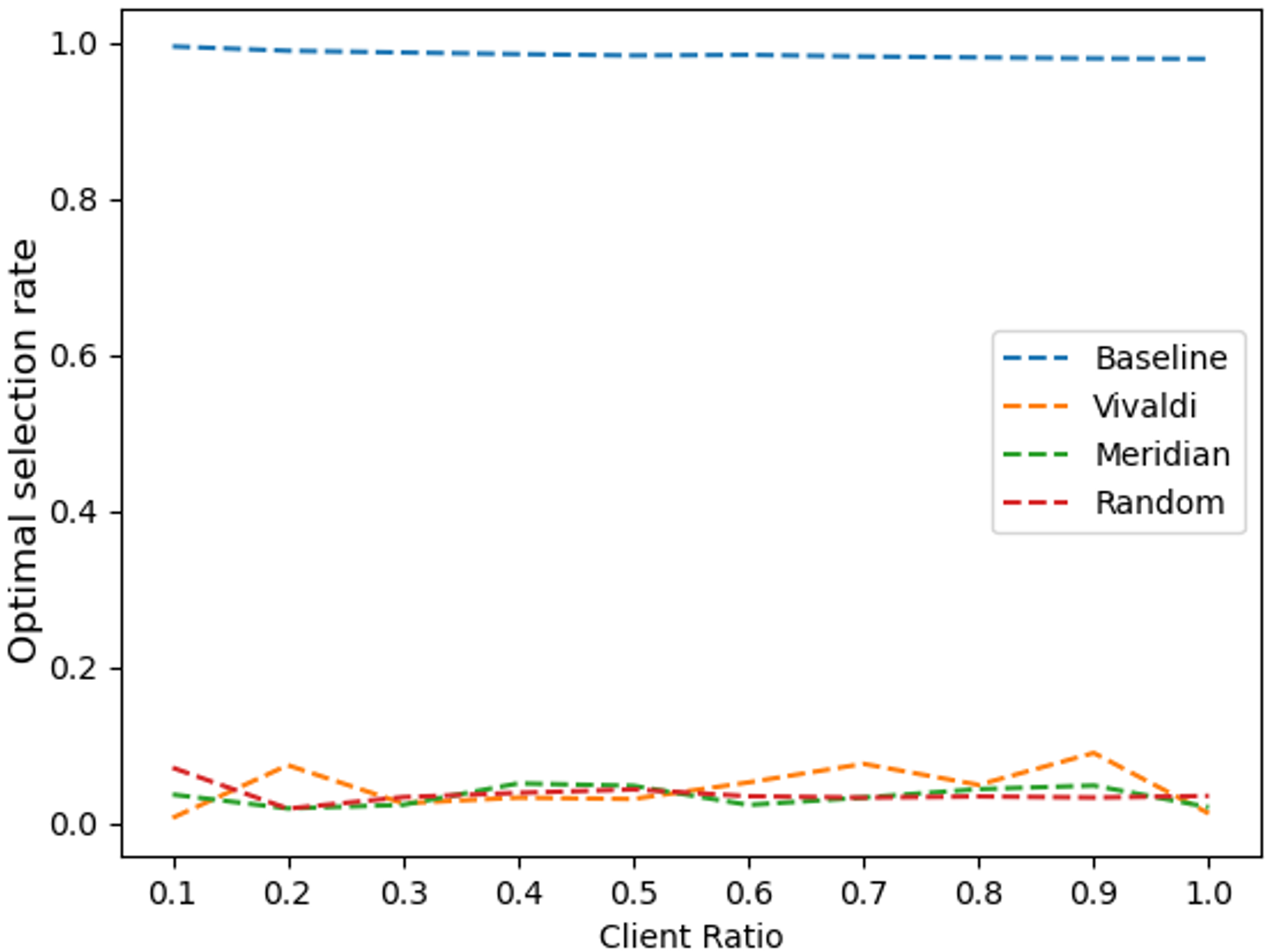}
        \subcaption{Optimal Selection Rate}
        \label{fig:servicediscovery:rate}
    \end{subfigure}
    \hfill
    \begin{subfigure}{0.32\linewidth}
        \centering
        \includegraphics[width=\linewidth]{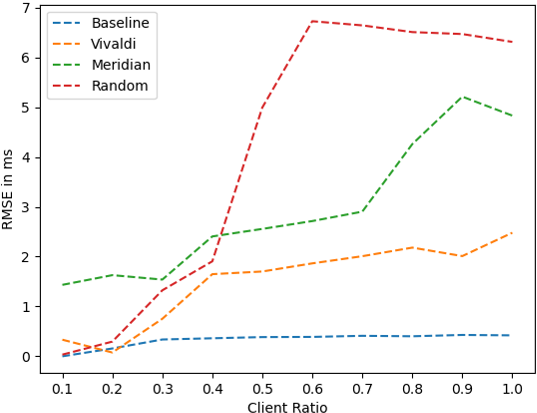}
        \subcaption{Selection Error}
        \label{fig:servicediscovery:rmse}
    \end{subfigure}
    \hfill
    \begin{subfigure}{0.32\linewidth}
        \centering
        \includegraphics[width=\linewidth]{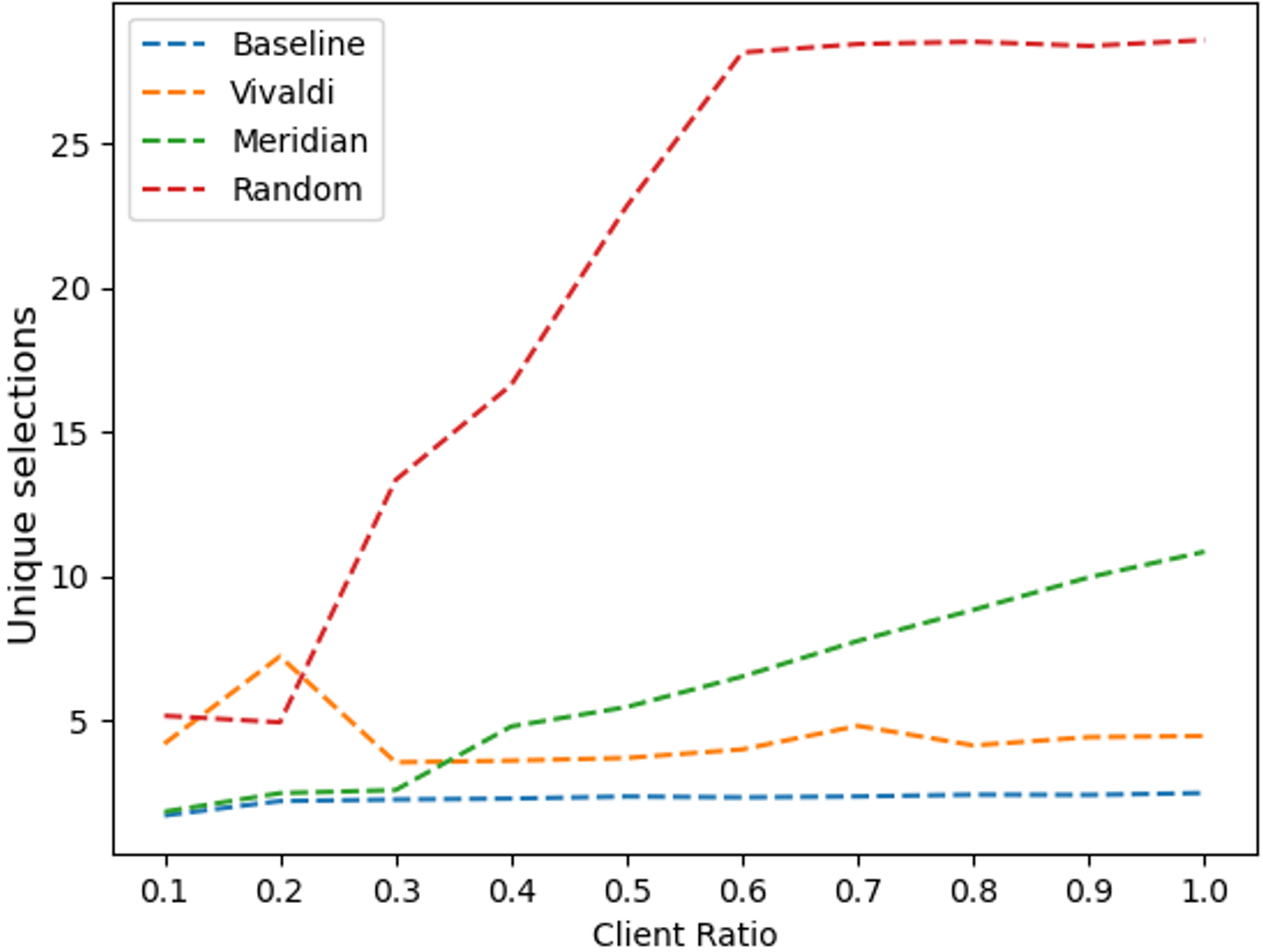}
        \subcaption{Unique Selections}
        \label{fig:servicediscovery:unique}
    \end{subfigure}
    \caption{The rate of optimal node selection (\cref{fig:servicediscovery:rate}) is 100\% for the baseline but less than 10\% for all other techniques, including the random approach. Nevertheless, Meridian and Vivaldi show a low RMSE selection error in terms of additional network delay over the optimal node selection (\cref{fig:servicediscovery:rmse}). The higher number of unique selections for NCS over the baseline (\cref{fig:servicediscovery:unique}) also leads to better load balancing across nodes.}
    \label{fig:servicediscovery}
\end{figure*}

We show the performance of service selection \cref{fig:servicediscovery}.
Compared to the baseline, all NCS perform poorly when it comes to selecting the optimal node, as shown in \cref{fig:servicediscovery:rate}.
We see in \cref{fig:servicediscovery:rmse}, however, that both NCS outperform the random approach in selection error, especially with higher client ratios.
Vivaldi outperforms Meridian in this metric.
While they mostly cannot find the \emph{optimal} node, both NCS find \emph{good} nodes.

\cref{fig:servicediscovery:unique} shows the number of unique selections between fog nodes.
As expected, the random approach chooses the most fog nodes as it just randomly forwards nodes from the network to the client.
Again, Vivaldi and Meridian are somewhere in between the two reference systems.
While the Vivaldi system has more unique selections in the lower client ratios, the Meridian system increases its unique selections with an increasing client ratio.

\subsubsection*{Network Latency}
\label{sec:analysis:latency}

\begin{figure*}
    \begin{subfigure}{0.49\linewidth}
        \centering
        \includegraphics[width=\linewidth]{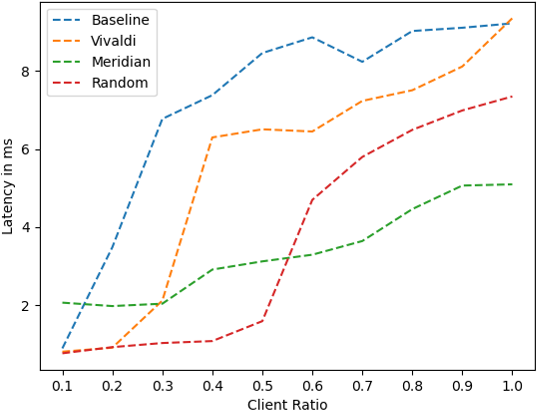}
        \subcaption{Mean Latency}
        \label{fig:clientconnection:meanlatency}
    \end{subfigure}
    \hfill
    \begin{subfigure}{0.49\linewidth}
        \centering
        \includegraphics[width=\linewidth]{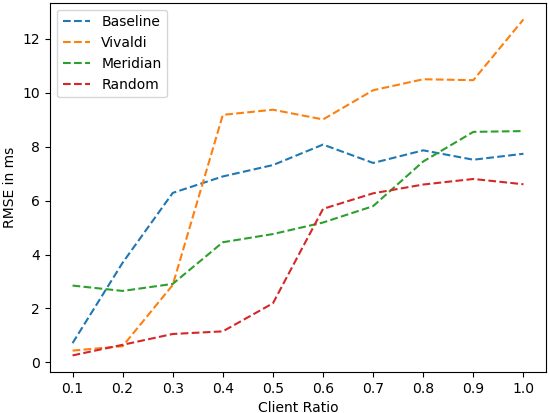}
        \subcaption{Connection Error}
        \label{fig:clientconnection:rttrmse}
    \end{subfigure}
    \caption{Actual client access latency also depends on available slots and available bandwidth, and surprisingly the mean latency results (\cref{fig:clientconnection:meanlatency}) show the random approach outperforming others with low client ratios. At higher client ratios, Meridian is best. Both approaches are good at load balancing, reducing additional delay caused by reconnection and limited bandwidth.}
    \label{fig:clientconnection}
\end{figure*}

\cref{fig:clientconnection} show the communication latency from clients to their identified fog nodes in different NCS approaches.
The mean latency (\cref{fig:clientconnection:meanlatency}) increases with the client ratio because of the increased load on the fog platform and the decreased bandwidth available per client.
The baseline and Vivaldi feature a higher average latency per client compared to Meridian and the random approach, which can be explained by the uniqueness of node selection.
Because the baseline and Vivaldi do not distribute the load of the clients over the entire network, the bandwidth of these nodes decreases.
This limited bandwidth translates into a higher transmission and queueing delay.
By not accounting for fog nodes with less capacity the random approach overloads these nodes, leading to increased latency.
Only the Meridian system is not completely agnostic of the resources of the fog nodes as fully loaded nodes do not take part in the selection process.

\cref{fig:clientconnection:rttrmse} shows the error between achieved RTT and theoretically optimal RTT.
Surprisingly, we can observe that the Vivaldi system has the highest connection error, especially in the higher client ratios.
The other systems deviate widely from each other in the lower client ratios but converge to a similar error with an increasing client ratio.

\section{Conclusion}
\label{sec:conclusion}

The results of our simulation show that NCS are powerful techniques to find nearest nodes in a distributed system that consider real network distance instead of, e.g., geographical location.
Surprisingly, however, we find that identifying the closest node is often not of upmost concern to clients in a fog network as dense as that of our simulation:
In most cases, the additional latency of suboptimal nodes was negligible.
Instead, we find the major driver for the efficiency of a node selection algorithm to be how well it balances load across fog nodes.
For future research on fog platforms we thus suggest focussing on distributed load balancing within bounds of service-level objectives.

\begin{acks}
    Supported by the \grantsponsor{DFG}{Deutsche Forschungsgemeinschaft (DFG, German Research Foundation)}{https://www.dfg.de/en/} -- \grantnum{DFG}{415899119}.
\end{acks}

\balance

\bibliographystyle{ACM-Reference-Format}
\bibliography{bibliography.bib}

\end{document}